# Duality Between Channel Capacity and Rate Distortion With Two-Sided State Information

Thomas M. Cover, *Fellow, IEEE,* and Mung Chiang, *Student Member, IEEE*

*Invited Paper*

*Abstract*—We show that the duality between channel capacity and data compression is retained when state information is available to the sender, to the receiver, to both, or to neither. We present a unified theory for eight special cases of channel capacity and rate distortion with state information, which also extends existing results to arbitrary pairs of independent and identically distributed (i.i.d.) correlated state information $(S_1, S_2)$ available at the sender and at the receiver, respectively. In particular, the resulting general formula for channel capacity $C = \max_{p(u,x|s_1)} [I(U; S_2, Y) - I(U; S_1)]$ assumes the same form as the generalized Wyner–Ziv rate distortion function $R(D) = \min_{p(u|x,s_1) p(\hat{x}|u,s_2)} [I(U; S_1, X) - I(U; S_2)]$.

*Index Terms*—Channel with state information, duality, multiuser information theory, rate distortion with state information, Shannon theory, writing on dirty paper.

## I. INTRODUCTION

SHANNON [16] remarked in his landmark paper on rate distortion:

"There is a curious and provocative duality between the properties of a source with a distortion measure and those of a channel. This duality is enhanced if we consider channels in which there is a cost associated with the different input letters … It can be shown readily that this [capacity cost] function is concave downward. Solving this problem corresponds, in a sense, to finding a source that is right for the channel and the desired cost … In a somewhat dual way, evaluating the rate distortion function for a source … the solution leads to a function which is convex downward. Solving this problem corresponds to finding a channel that is just right for the source and allowed distortion level."

Thus, the two fundamental limits of data communication and data compression are dual. Channel capacity is the maximum data transmission rate across a communication channel with the probability of decoding error approaching zero, and the rate distortion function is the minimum rate needed to describe a source under a distortion constraint.

In this paper, we look at four problems in data transmission in the presence of state information and four problems in data compression with state information. These data transmission and compression models have found applications in wireless communications where the fading coefficient is the state information at the sender, in capacity calculation for defective memory where the defective memory cell is the state information at the encoder, in digital watermarking where the original image is the state information at the sender, and in high-definition television (HDTV) systems where the noisy analog version of the TV signal is the state information at the decoder.

The eight problems in data transmission and data compression with state information will be seen to have similar answers with two odd exceptions. However, by putting all eight answers in a common form, we exhibit the duality between channel capacity and rate distortion theory in the presence of state information. In the process, we are led to a single more general theorem covering capacity and data compression with different state information at the source and at the destination. Surprisingly, it turns out that the unifying formula is the odd exception, which can be traced back to the Wyner–Ziv formula for rate distortion with state information at the receiver and to the counterpart Gelfand–Pinsker capacity formula for channels with state information at the sender.

## II. A CLASS OF CHANNELS WITH STATE INFORMATION

Recall that for a channel $\{\mathcal{X}, p(y|x), \mathcal{Y}\}$ without state information, the rate $R$ is achievable if there is a sequence of $(2^{nR}, n)$ codes with encoder $X^n: \{1, 2, \ldots 2^{nR}\} \to \mathcal{X}^n$ and decoder $\hat{W}: \mathcal{Y}^n \to \{1, 2, \ldots, 2^{nR}\}$, such that

$$P_e = \frac{1}{M} \sum_{i=1}^{M} \Pr\{\hat{W}(Y^n) \neq W | W = i\} \to 0, \qquad \text{as } n \to \infty$$

where $M = 2^{nR}$. The channel capacity is the supremum of achievable rates.

A class of discrete memoryless channels $\{\mathcal{X}, p(y|x, s), \mathcal{Y}, \mathcal{S}\}$ with state information $S^n$, $S_i$ independent and identically distributed (i.i.d.) $\sim p(s)$, has been studied by several groups of researchers, including Kusnetsov and Tsybakov [12], Gelfand

Manuscript received June 13, 2001; revised December 18, 2001. The work of T. M. Cover was supported in part by the NSF under Grant CCR-9973134 and MURI DAAD-19-99-1-0215. The work of M. Chiang was supported by the Hertz Foundation Graduate Fellowship and the Stanford Graduate Fellowship. The material in this paper was presented in part at the IEEE International Symposium of Information Theory and its Applications, HI, November 2000, and at the IEEE International Symposium of Information Theory, Washington, DC, June 2001.

The authors are with the Electrical Engineering Department, Stanford University, Stanford, CA 94305 USA (e-mail: cover@ee.stanford.edu; chiangm@stanford.edu).

Communicated by S. Shamai, Guest Editor.

Publisher Item Identifier S 0018-9448(02)04028-2.

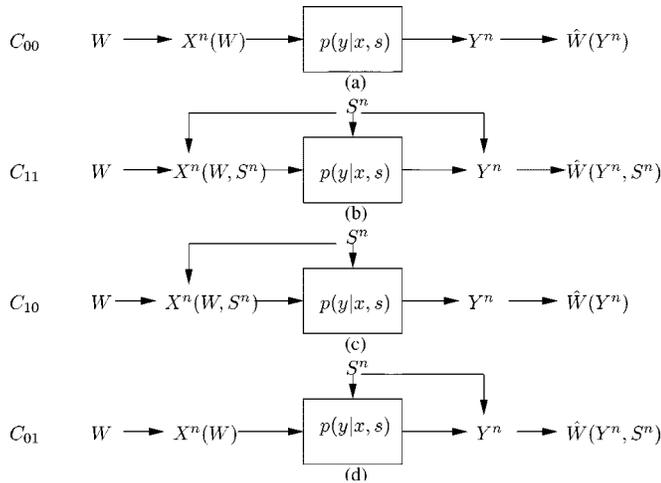

Fig. 1. Channels with state information. Four special cases. (a) $C_{00} = \max_{p(x)} I(X;Y)$. (b) $C_{11} = \max_{p(x|s)} I(X;Y|S)$. (c) $C_{10} = \max_{p(u,x|s)} [I(U;Y) - I(U;S)]$. (d) $C_{01} = \max_{p(x)} I(X;Y|S)$.

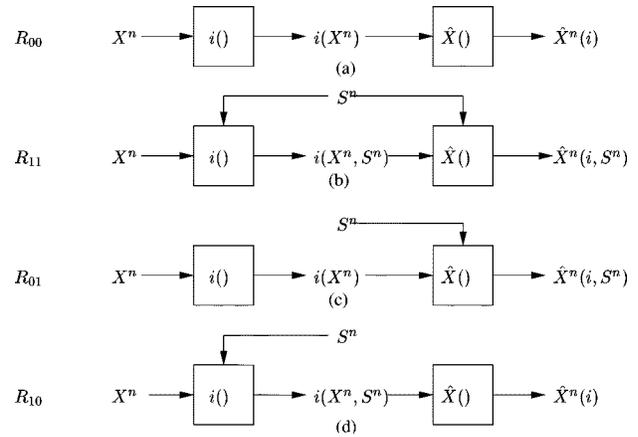

Fig. 2. Rate distortion with state information. Four special cases. (a) $R_{00} = \min_{p(\hat{x}|x)} I(X;\hat{X})$. (b) $R_{11} = \min_{p(\hat{x}|x,s)} I(X;\hat{X}|S)$. (c) $R_{01} = \min_{p(u|x)p(\hat{x}|u,s)} [I(U;X) - I(U;S)]$. (d) $R_{10} = \min_{p(\hat{x}|x)} I(X;\hat{X})$.

and Pinsker [11], and Heegard and El Gamal [10]. The state $S^n = (S_1, S_2, \ldots, S_n)$ is available noncausally. Fig. 1 shows the four special cases of channels with noncausal state information.

As the first case, we denote by $C_{00}$ the channel capacity when neither the sender nor the receiver knows the state information $S^n$. As in the rest of the paper, the first subscript under $C$ and $R$ denotes the availability of state information to the sender, and the second subscript the availability of state information to the receiver. When state information is not available to either the sender or the receiver, the channel capacity is the same as the capacity for the channel without state information. Therefore,

$$C_{00} = \max_{p(x)} I(X;Y). \quad (1)$$

Similarly, we denote by $C_{11}$ the channel capacity when both the sender and the receiver know the state information, where the encoder maps $(W, S^n)$ to $X^n$ and the decoder maps $(Y^n, S^n)$ to $\{1, 2, \ldots, 2^{nR}\}$. Here the channel capacity is

$$C_{11} = \max_{p(x|s)} I(X;Y|S). \quad (2)$$

This is achieved by finding the channel capacity for each state and using the corresponding capacity-achieving code on the subsequence of times where the state $S_i$ takes on a given value.

We denote by $C_{01}$ the capacity when only the receiver knows $S^n$. While the mutual information to be maximized is still conditioned on state $S$, we only maximize over $p(x)$, rather than $p(x|s)$, since state information is no longer available to the sender. The channel capacity was proved in [10] to be

$$C_{01} = \max_{p(x)} I(X;Y|S). \quad (3)$$

As the fourth case, we denote by $C_{10}$ the capacity when only the sender knows $S^n$. Thus, the encoding of a message $W \in 2^{nR}$ is given by $X^n(W, S^n)$ and the decoding by $\hat{W}(Y^n)$. The capacity has been established by Gelfand and Pinsker [11] to be

$$C_{10} = \max_{p(u,x|s)} [I(U;Y) - I(U;S)]. \quad (4)$$

It is particularly interesting to observe that an auxiliary random variable $U$ is needed to express capacity when state information is available only to the sender. The implications will be discussed in the following sections, and the odd form of (4) will be shown to be the fundamental form. We will also show that all four channel capacities look like (4) and are simple corollaries of Theorem 1 in Section VI.

### III. RATE DISTORTION WITH STATE INFORMATION

We now turn to rate distortion theory. Let $\{(X_k, S_k)\}$ i.i.d. $\sim p(x, s)$ be a sequence of independent drawings of jointly distributed random variables $X$ and $S$. We are given a distortion measure $d(x, \hat{x})$. We wish to describe $\{X_k\}$ at rate $R$ bits per symbol and reconstruct $\{\hat{X}_k\}$ with distortion $D$. The sequence $\{X_k\}$ is encoded in blocks of length $n$ into a binary stream of rate $R$, which will in turn be decoded as a sequence $\{\hat{X}_k\}$ in the reproduction alphabet. The average distortion is

$$D = \frac{1}{n} \sum_{k=1}^{n} E[d(X_k, \hat{X}_k)].$$

We say that rate $R$ is achievable at distortion level $D$ if there exists a sequence of $(2^{nR}, n)$ codes

$$i : \mathcal{X}^n \to \{1, 2, \ldots, 2^{nR}\}$$
$$\hat{X}^n : \{1, 2, \ldots, 2^{nR}\} \to \hat{\mathcal{X}}^n$$

such that $Ed(X^n, \hat{X}^n(i(X^n))) \leq D$. The rate distortion function $R(D)$ is the infimum of the achievable rates with distortion $D$. Fig. 2 shows the four special cases of rate distortion with state information.

The rate necessary to achieve distortion $D$ will be denoted by $R_{00}(D)$, or simply $R_{00}$, if state information $S^n$ is available to neither the encoder nor the decoder. This is the same problem as the standard rate distortion problem without state information, and the rate distortion function is given by

$$R_{00}(D) = \min_{p(\hat{x}|x)} I(X;\hat{X}) \quad (5)$$

where the minimum is over all $p(\hat{x}|x)$ such that

$$\sum_{x,\hat{x}} p(x) p(\hat{x}|x) d(x, \hat{x}) \leq D.$$

When the state $S^n$ is available to both the encoder and the decoder, we denote the rate-distortion function by $R_{11}$. Here, a $(2^{nR}, n)$ code consists of an encoding map $i: \mathcal{X}^n \times \mathcal{S}^n \to \{1, 2, \ldots, 2^{nR}\}$ and a reconstruction map $\hat{X}^n: \{1, 2, \ldots, 2^{nR}\} \times \mathcal{S}^n \to \hat{\mathcal{X}}^n$. Thus, both $i(\cdot)$ and $\hat{X}(\cdot)$ depend on the state $S^n$. In this case, both the mutual information to be minimized and the probability distribution of $\hat{X}$ are now conditioned on $S$. Thus,

$$R_{11}(D) = \min_{p(\hat{x}|x,s)} I(X; \hat{X}|S) \qquad (6)$$

where the minimum is taken over all $p(\hat{x}|x, s)$ such that

$$\sum_{x,\hat{x},s} p(x, s) p(\hat{x}|x, s) d(x, \hat{x}) \leq D.$$

We denote by $R_{10}$ the rate distortion when only the encoder knows the state information. It was shown by Berger [1] that

$$R_{10}(D) = \min_{p(\hat{x}|x)} I(X; \hat{X}) \qquad (7)$$

where the minimum is taken over all $p(\hat{x}|x)$ such that

$$\sum_{x,\hat{x}} p(x) p(\hat{x}|x) d(x, \hat{x}) \leq D.$$

Finally, let $R_{01}$ be the rate-distortion function with state information $S^n$ at the decoder, as shown in Fig. 2(c). This is a much more difficult problem. Wyner and Ziv [19] proved the rate distortion function to be

$$R_{01}(D) = \min_{p(u|x)p(\hat{x}|u,s)} [I(U; X) - I(U; S)] \qquad (8)$$

where the minimum is over all $p(u|x)p(\hat{x}|u, s)$ such that

$$\sum_{x,\hat{x},u,s} p(x, s) p(u|x) p(\hat{x}|u, s) d(x, \hat{x}) \leq D$$

and $|\mathcal{U}| \leq |\mathcal{X}| + 1$.

Comparing Fig. 1 with Fig. 2, it is evident that the setups of the channel capacity and rate-distortion problems are dual. We will show that all results reviewed in this section are simple corollaries of a general result on rate distortion with state information proved in Theorem 2 in Section VI.

## IV. DUALITY BETWEEN RATE DISTORTION AND CHANNEL CAPACITY

We first investigate the duality and equivalence relationships of these channel capacity and rate distortion problems with state information. With the following transformation it is easy to verify that (1), (2), and (4) are dual to (5), (6), and (8), respectively. The left column corresponds to channel capacity and the right column to rate distortion.

*Transformation:* Correspondence between channel capacities in Fig. 1 and rate distortions in Fig. 2:

$$C \leftrightarrow R(D) \qquad (9)$$

$$\text{maximization} \leftrightarrow \text{minimization} \qquad (10)$$

$$C_{00} \leftrightarrow R_{00} \qquad (11)$$

$$C_{11} \leftrightarrow R_{11} \qquad (12)$$

$$C_{10} \leftrightarrow R_{01} \qquad (13)$$

$$Y(\text{received symbol}) \leftrightarrow X(\text{source}) \qquad (14)$$

$$X(\text{transmitted symbol}) \leftrightarrow \hat{X}(\text{estimation}) \qquad (15)$$

$$S(\text{state}) \leftrightarrow S(\text{state}) \qquad (16)$$

$$U(\text{auxiliary}) \leftrightarrow U(\text{auxiliary}). \qquad (17)$$

The duality is evident. In particular, the supremum of achievable data rates for a channel with state information at the encoder has a formula dual to the infimum of data rates describing a source with state information at the decoder. Note that the roles of the sender and the receiver in channel capacity are opposite to those of the encoder and the decoder in rate distortion, which is seen by exchanging the first and the second subscript of $C$ and $R$.

There is one particularly interesting symmetry between $R_{01}$ and $C_{10}$ in terms of the distribution over which the minimization and the maximization is taken. Consider the rate distortion function $R_{01}(D)$. We can first minimize over the probability distribution $p(u|x)$ and then minimize over deterministic functions $f(u, s) = \hat{x}$. Symmetrically, for the channel capacity $C_{10}$, we can restrict the maximization of $p(u, x|s)$ to a maximization over $p(u|s)$, followed by a maximization over deterministic functions $f(u, s) = x$. In both problems, restricting the extremization to deterministic functions incurs no loss of generality. This algebraic simplification of the $C_{10}$ and $R_{01}$ formulas is dual.

We pause here to comment on the meaning of duality. There are two common notions of duality: complementarity and isomorphism. The notion of good and bad is an example of complementarity, and the notion of inductance and capacitance is an example of isomorphism. Nicely enough, these two definitions of duality are complementary and are themselves dual.

Like duality in optimization theory, the information-theoretic duality relationships we consider in this paper include both complementarity and isomorphism. The roles of the encoder and the decoder are complementary. The minimization of the mutual information quantity in the rate distortion problem, which is a convex function, is complementary to the maximization of the mutual information quantity in the channel capacity problem, which is a concave function. Furthermore, complementing the encoder–decoder pair and the maximization–minimization pair makes the channel capacity formula isomorphic to the rate-distortion function. Such duality relationships are maintained when state random variables and auxiliary random variables are included in the models. These complementary and isomorphic duality relationships are further illuminated through the unification in Section VI of the eight special cases considered so far.

## V. RELATED WORK

We now give a brief nonexhaustive review of related work on state information and duality.

The duality between $C_{10}$ and $R_{01}$ has been noted by several researchers. It was pointed out by the authors in [4]–[6] in the study of communication over an unreliable channel. Duality for the Gaussian case was also pointed out in Chou, Pradhan, and Ramchandran [7] for distributed data compression and the associated coding methods. In digital watermarking schemes, both the Gaussian and binary cases of the duality between $C_{10}$ and $R_{01}$ were presented in [2], together with the associated coding methods. The geometric interpretation of this duality in the Gaussian case was developed in [17]. For channels with two-sided state information, Caire and Shamai [3] showed the optimal coding method when state information at the sender is a deterministic function of the state information at the receiver. In this paper, we show that there are duality relationships among all eight special cases of channel capacity and rate distortion with state information. These special cases are then unified into two-sided state information generalizations of the Wyner–Ziv and Gelfand–Pinsker formulas.

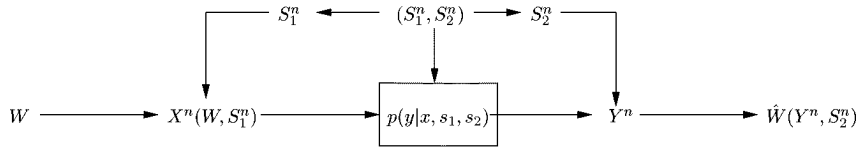

Fig. 3. Channel capacity with two-sided state information, where $(S_{1,i}, S_{2,i})$ are i.i.d. $\sim p(s_1, s_2)$: $C = \max_{p(u,x|s_1)} [I(U; S_2, Y) - I(U; S_1)]$.

We note that there is a different model of channel with state information proposed by Shannon [15]. Shannon studied discrete memoryless channels in which only causal state information is available, i.e., the input symbol $X_1(W, S^i)$ at time $i$ depends only on the past and present states, and not on the future $S_{i+1}^n$. The capacity is $C = \max I(X(\cdot); Y)$, where the maximization is over all distributions on functions $X(\cdot): \mathcal{S} \to \mathcal{X}$. The capacity for the Gaussian version of Shannon's channel is not known, though coding schemes have been proposed in [9]. This model is not expected to be dual to the rate-distortion problem, since there is an intrinsic noncausality in encoding and decoding of blocks $X^n$ in the rate-distortion problem. Therefore, only the noncausal version of channel capacity with state information corresponds naturally to rate distortion with state information.

From a coding viewpoint, the random binning proofs of $C_{10}$ and $R_{01}$ are also dual to each other and they resemble trellis-coded modulation and coset codes. In fact, shifted lattice codes or coset codes can be viewed as practical ways to implement the random binning and sphere covering ideas. Lattice codes were used for channels with state information at the encoder in [9], and for rate distortion with state information at the decoder in [21].

The increase and the decrease of channel capacity and rate distortion when state information is available have also been studied. The duality in the differences $C_{10} - C_{00}$ and $R_{01} - R_{00}$ was shown in [5] for the Gaussian case. The rate loss in the Wyner–Ziv problem was shown to be bounded by a constant in [20].

An input cost function can be introduced to make channel capacity with state information more closely resemble rate distortion with state information. Examples for point-to-point channels without state information can be found in [14]. Models with input cost have not been studied for channels with state information. For rate distortion with state information, a state-information-dependent distortion measure was studied in [13]. In particular, the asymmetry between $C_{01} \neq C_{00}$ and $R_{10} = R_{00}$ can be resolved by introducing a state-dependent distortion measure.

The general problem of the tradeoff between state estimation and channel capacity was developed in [18], where the receiver balances two objectives: decoding the message and estimating the state. The sender, knowing the state information, can choose between maximizing the throughput of the intended message and helping the receiver estimate the channel state with the smallest estimation error.

## VI. CHANNEL CAPACITY AND RATE DISTORTION WITH TWO-SIDED STATE INFORMATION

The results for $C_{00}$, $C_{10}$, $C_{01}$, $C_{11}$, and correspondingly $R_{00}$, $R_{10}$, $R_{01}$, $R_{11}$, assume different forms. We wish to put these results in a common framework. Despite the less straightforward form of $C_{10}$ and $R_{01}$, we proceed to show that all the other cases can best be expressed in that form. Thus, the Gelfand–Pinsker and the Wyner–Ziv formulas are the fundamental forms of channel capacity and rate distortion with state information, rather than the orphans. We also consider a generalization where the sender and the receiver have correlated but different state information.

First consider channel capacity. We assume that the channel is embedded in some environment with state information $S_1^n$ available to the sender, correlated state information $S_2^n$ available to the receiver, and a memoryless channel with transition probability $p(y|x, s_1, s_2)$ that depends on the input $X$ and the state $(S_1, S_2)$ of the environment. We assume that $(S_{1,i}, S_{2,i})$ are i.i.d. $\sim p(s_1, s_2)$, $i = 1, 2, \ldots$. The output $Y^n$ has conditional distribution

$$p(y^n|x^n, s_1^n, s_2^n) = \prod_{i=1}^n p(y_i|x_i, s_{1,i}, s_{2,i}).$$

The encoder $X^n(W, S_1^n), W \in 2^{nR}$, and the decoder $\hat{W}(Y^n, S_2^n)$ defining a $(2^{nR}, n)$ code are shown in Fig. 3. The resulting probability of error is $P_e^n = \Pr\{\hat{W}(Y^n, S_2^n) \neq W\}$, where $W$ is drawn according to a uniform distribution over $\{1, 2, \ldots, 2^{nR}\}$. A rate $R$ is achievable if there exists a sequence of $(2^{nR}, n)$ codes with $P_e^n \to 0$. The capacity $C$ is the supremum of the achievable rates.

*Theorem 1:* The memoryless channel $p(y|x, s_1, s_2)$ with state information $(S_{1,i}, S_{2,i})$ i.i.d. $\sim p(s_1, s_2)$, with $S_1^n$ available to the sender and $S_2^n$ available to the receiver, has capacity

$$C = \max_{p(u,x|s_1)} [I(U; S_2, Y) - I(U; S_1)]. \qquad (18)$$

*Proof:* Section VII contains the proof.

*Corollary 1:* The four capacities $C_{00}, C_{01}, C_{10}, C_{11}$ with state information, given in (1)–(4), are special cases of Theorem 1.

*Proof:*

Case 1: No state information: $S_1 = \phi, S_2 = \phi$. Here, $(S_2, Y) = Y$, $I(U; S_1) = 0$, and Theorem 1 reduces to

$$C = \max_{p(u,x|s_1)} [I(U; S_2, Y) - I(U; S_1)] \qquad (19)$$

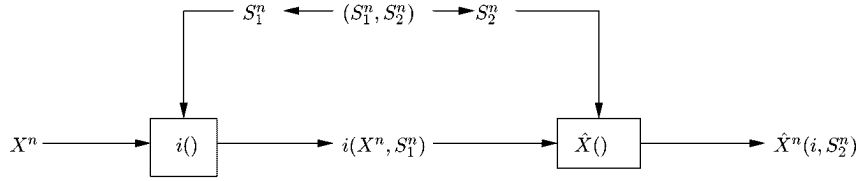

Fig. 4. Rate distortion with two-sided state information, where $(X_i, S_{1,i}, S_{2,i})$ are i.i.d. $\sim p(x, s_1, s_2)$: $R(D) = \min_{p(u|x,s_1)p(\hat{x}|u,s_2)} [I(U; S_1, X) - I(U; S_2)]$.

$$= \max_{p(u,x)} I(U; Y) \quad (20)$$

$$= \max_{p(x)} I(X; Y) \quad (21)$$

$$= C_{00} \quad (22)$$

where we use the fact that under the allowed distribution $p(s)p(u)p(x|u)p(y|x, s)$, $U \to X \to Y$ forms a Markov chain. Therefore, by the data processing inequality

$$\max_{p(u,x)} I(U; Y) \leq \max_{p(x)} I(X; Y)$$

with equality iff $U = X$, and equation (21) follows.

Case 2: State information at the receiver: $S_1 = \phi$, $S_2 = S$. Here, $I(U; S_1) = 0$, and Theorem 1 reduces to

$$C = \max_{p(u,x|s_1)} [I(U; S_2, Y) - I(U; S_1)] \quad (23)$$

$$= \max_{p(u,x)} I(U; S, Y) \quad (24)$$

$$= \max_{p(u,x)} [I(U; S) + I(U; Y|S)] \quad (25)$$

$$= \max_{p(u,x)} I(U; Y|S) \quad (26)$$

$$= \max_{p(x)} I(X; Y|S) \quad (27)$$

$$= C_{01} \quad (28)$$

where we have used the fact that $U$ and $S$ are independent under the allowed distribution $p(s)p(u)p(x|u)p(y|x, s)$). Also, under the allowed distribution, $U \to X \to Y$ forms a Markov chain conditioned on $S$. Therefore, by the conditional data processing inequality, $\max_{p(u,x)} I(U; Y|S) \leq \max_{p(x)} I(X; Y|S)$, with equality iff $U = X$, and (27) follows.

Case 3: (The Gelfand–Pinsker formula) State information at the sender: $S_1 = S$, $S_2 = \phi$. Here $(S_2, Y) = Y$, and Theorem 1 reduces to

$$C = \max_{p(u,x|s_1)} [I(U; S_2, Y) - I(U; S_1)] \quad (29)$$

$$= \max_{p(u,x|s)} [I(U; Y) - I(U; S)] \quad (30)$$

$$= C_{10}. \quad (31)$$

Case 4: State information at the sender and the receiver: $S_1 = S$, $S_2 = S$, and Theorem 1 reduces to

$$C = \max_{p(u,x|s_1)} [I(U; S_2, Y) - I(U; S_1)] \quad (32)$$

$$= \max_{p(u,x|s)} [I(U; S, Y) - I(U; S)] \quad (33)$$

$$= \max_{p(u,x|s)} [I(U; S) + I(U; Y|S) - I(U; S)] \quad (34)$$

$$= \max_{p(x|s)} I(X; Y|S) \quad (35)$$

$$= C_{11} \quad (36)$$

where we have used the fact that under the allowed distribution $p(s)p(u|s)p(x|u, s)p(y|x, s)$, $U \to X \to Y$ forms a Markov chain conditioned on $S$. Therefore, by the conditional data processing inequality

$$\max_{p(u,x|s)} I(U; Y|S) \leq \max_{p(x|s)} I(X; Y|S)$$

with equality iff $U = X$, and equation (35) follows.

We now find the rate distortion function for the general problem depicted in Fig. 4, where $i: \mathcal{X}^n \times \mathcal{S}_1^n \to 2^{nR}$, $\hat{X}: 2^{nR} \times \mathcal{S}_2^n \to \hat{\mathcal{X}}^n$

$$Ed(X^n, \hat{X}^n(i(X^n, S_1^n), S_2^n)) = D$$

and $(S_{1,i}, S_{2,i})$ i.i.d. $\sim p(s_1, s_2)$. Let $R(D)$ be the minimum achievable rate with distortion $D$.

*Theorem 2:* For a bounded distortion measure $d(x, \hat{x})$ and $(X_i, S_{1,i}, S_{2,i})$ i.i.d. $\sim p(x, s_1, s_2)$, where $\mathcal{X}, \mathcal{S}_1, \mathcal{S}_2$ are finite sets, let $S_1^n$ be available to the encoder and $S_2^n$ to the decoder. The rate distortion function is

$$R(D) = \min_{p(u|x,s_1)p(\hat{x}|u,s_2)} [I(U; S_1, X) - I(U; S_2)] \quad (37)$$

where the minimization is under the distortion constraint

$$\sum_{x, u, s_1, s_2, \hat{x}} d(x, \hat{x}) p(x, s_1, s_2) p(u|x, s_1) p(\hat{x}|u, s_2) \leq D.$$

*Proof:* Section VIII contains the proof.

*Corollary 2:* The four rate distortion functions $R_{00}(D)$, $R_{01}(D)$, $R_{10}(D)$, $R_{11}(D)$ with state information, given in (5)–(8), are special cases of Theorem 2.

*Proof:* We evaluate

$$R(D) = \min_{p(u|x,s_1)p(\hat{x}|u,s_2)} [I(U; S_1, X) - I(U; S_2)]$$

under the distortion constraint $Ed(X^n, \hat{X}^n) \leq D$ in each of the four cases.

Case 1: No state information: $S_1 = \phi$, $S_2 = \phi$. Here, $(S_1, X) = X$ and $I(U; S_2) = 0$, and Theorem 2 reduces to

$$R(D) = \min_{p(u|x,s_1)p(\hat{x}|u,s_2)} [I(U; S_1, X) - I(U; S_2)] \quad (38)$$

$$= \min_{p(u|x)p(\hat{x}|u)} I(U; X) \quad (39)$$

$$= \min_{p(\hat{x}|x)} I(\hat{X}; X) \quad (40)$$

$$= R_{00}(D) \quad (41)$$

where we have used the fact that, under the allowed minimizing distribution $p(x, u, \hat{x}) = p(x)p(u|x)p(\hat{x}|u)$, $X \to U \to \hat{X}$ forms a Markov chain. Therefore, by the data processing inequality

$$\min_{p(u,\hat{x}|x)} I(U; X) \geq \min_{p(\hat{x}|x)} I(\hat{X}; X)$$

with equality iff $U = \hat{X}$, and equation (40) follows.

Case 2: (The Wyner–Ziv formula) State information at the receiver: $S_1 = \phi$, $S_2 = S$. Here, $(S_1, X) = X$, and Theorem 2 reduces to

$$R(D) = \min_{p(u|x,s_1)p(\hat{x}|u,s_2)} [I(U; S_1, X) - I(U; S_2)] \quad (42)$$

$$= \min_{p(u|x)p(\hat{x}|u,s)} [I(U; X) - I(U; S)] \quad (43)$$

$$= R_{01}(D). \quad (44)$$

Case 3: State information at the sender: $S_1 = S$, $S_2 = \phi$. Here, $I(U; S_2) = 0$, and Theorem 2 reduces to

$$R(D) = \min_{p(u|x,s_1)p(\hat{x}|u,s_2)} [I(U; S_1, X) - I(U; S_2)] \quad (45)$$

$$= \min_{p(u|x,s)p(\hat{x}|u)} I(U; S, X) \quad (46)$$

$$= \min_{p(u|x,s)p(\hat{x}|u)} [I(U; X) + I(U; S|X)] \quad (47)$$

$$\geq \min_{p(u|x,s)p(\hat{x}|u)} I(U; X) \quad (48)$$

$$= \min_{p(\hat{x}|x,s)} I(X; \hat{X}) \quad (49)$$

$$= \min_{p(\hat{x}|x)} I(X; \hat{X}) \quad (50)$$

where (48) is due to $I(U; S|X) \geq 0$ and (49) is due to the data-processing inequality on the Markov chain $X \to U \to \hat{X}$ with equality iff $U = \hat{X}$. Equation (50) holds because letting $p(\hat{x}|x, s) = p(\hat{x}|x)$ does not change the functionals $I(X; \hat{X})$ and $Ed(X, \hat{X})$. Since $U = \hat{X}$ and $p(\hat{x}|x, s) = p(\hat{x}|x)$, we have $I(U; S|X) = 0$ and inequality (48) is achieved with equality. Therefore, we have the desired reduction

$$R(D) = \min_{p(\hat{x}|x)} I(X; \hat{X}) \quad (51)$$

$$= R_{10}(D). \quad (52)$$

Case 4: State information at the sender and the receiver: $S_1 = S$, $S_2 = S$. Theorem 2 reduces to

$$R(D) = \min_{p(u|x,s_1)p(\hat{x}|u,s_2)} [I(U; S_1, X) - I(U; S_2)] \quad (53)$$

$$= \min_{p(u|x,s)p(\hat{x}|u,s)} [I(U; S, X) - I(U; S)] \quad (54)$$

$$= \min_{p(u|x,s)p(\hat{x}|u,s)} [I(U; S) + I(U; X|S) - I(U; S)] \quad (55)$$

$$= \min_{p(\hat{x}|x,s)} I(\hat{X}; X|S) \quad (56)$$

$$= R_{11}(D) \quad (57)$$

where we have used the fact that under the allowed minimizing distribution $p(x|s)p(u|x, s)p(\hat{x}|u, s)$, $X \to U \to \hat{X}$ forms a Markov chain conditioned on $S$. Therefore, by the conditional data processing inequality

$$\min_{p(u,\hat{x}|x,s)} I(U; X|S) \geq \min_{p(\hat{x}|x,s)} I(\hat{X}; X|S)$$

with equality iff $U = \hat{X}$, and (56) follows. This concludes the proof.

The general results in Theorems 1 and 2 are dual and assume the form of the Gelfand–Pinsker and Wyner–Ziv formulas, respectively. In particular, the roles of $S_1$ and $S_2$ in channel capacity and rate distortion are dual. The corresponding Corollarys 1 and 2 yield the eight special cases. Notice that the apparent asymmetry between $C_{01}$ and $R_{10}$ is resolved in this unification.

## VII. PROOF OF THEOREM 1

The proof in this section closely follows the proof in [11].

We must prove that the capacity of a discrete memoryless channel $p(y|x, s_1, s_2)$ with two-sided state information is given by

$$C = \max_{p(u,x|s_1)} [I(U; S_2, Y) - I(U; S_1)].$$

We prove this under the condition that the alphabets $|\mathcal{X}|$, $|\mathcal{S}_1|$, $|\mathcal{S}_1|$, $|\mathcal{Y}|$ are finite.

We first give some remarks about how to achieve capacity. The main idea is to transfer the information conveying role of the channel input $X^n$ to some fictitious input $U^n$, so that the channel behaves like a discrete memoryless channel $U \to (Y, S_2)$ with

$$p(u^n)p(s_1^n, s_2^n, x^n | u^n)p(y^n | x^n, s_1^n, s_2^n)$$
$$= \prod_{i=1}^{n} p(u_i)p(s_{1,i}, s_{2,i}, x_i | u_i)p(y_i | x_i, s_{1,i}, s_{2,i}).$$

The capacity of this new channel is $I(U; Y, S_2)$. This can be achieved in the sense that $2^{nI(U;Y,S_2)}$ $U^n(i)$'s can be distinguished by the receiver $(Y^n, S_2^n)$. But there is a cost for setting up such a dependence; only a fraction $2^{-nI(U;S_1)}$ of the possible codewords $U^n(i)$ are jointly typical with $S_1^n$. Thus, only $2^{nI(U;Y,S_2)-nI(U;S_1)}$ distinguishable codewords $U^n(i)$ are available for transmission, those with the required empirical joint distribution on $(U^n, X^n, S_1^n)$.

Here, the transmission sequence $X^n$ is chosen so that $(U^n(i), X^n, S_1^n)$ is strongly jointly typical. Any randomly drawn $X^n \sim \prod_{j=1}^{n} p(x_j | u_j(i), s_{1,j})$ will do, but at capacity, this conditional distribution will be degenerate, and there will be only a negligible number (actually $2^{n\epsilon}$) of such conditionally typical $X^n$. Thus, $X^n$ will turn out to be a function of $(U^n(i), S_1^n)$, designed to make $(U^n(i), X^n, S_1^n)$ typical and to make the channel operate as a discrete memoryless channel from $U^n$ to $(Y^n, S_2^n)$.

We prove Theorem 1 in two parts. First, we show that

$$C = \max_{p(u,x|s_1)} [I(U; S_2, Y) - I(U; S_1)]$$

is achievable.

Let $p(u, x|s_1)$ be chosen to yield $C$. We are given that $(S_{1,i}, S_{2,i})$ are i.i.d.$\sim p(s_1, s_2)$, the encoder produces $X^n(W, S_1^n)$, and the decoder produces $\hat{W}(Y^n, S_2^n)$. The channel is memoryless

$$p(y^n|x^n, s_1^n, s_2^n) = \prod_i p(y_i|s_{1,i}, s_{2,i}, x_i).$$

Now consider a $(2^{nR}, n)$ code with encoder

$$X^n: \{1, 2, \ldots, 2^{nR}\} \times \mathcal{S}_1^n \to \mathcal{X}^n$$

and decoder

$$\hat{W}: \mathcal{Y}^n \times \mathcal{S}_2^n \to \{1, 2, \ldots, 2^{nR}\}.$$

The average probability of error $P_e^n$ is defined to be

$$P_e^n = \frac{1}{M} \sum_{i=1}^M \Pr\{\hat{W}(Y^n, S_2^n) \neq i | W = i\}.$$

The encoding and decoding strategy is as follows. First, generate $2^{n(I(U;Y,S_2)-2\epsilon)}$ i.i.d. sequences

$$U^n(i), i \in 2^{n(I(U;Y,S_2)-2\epsilon)}$$

according to distribution $\prod_{l=1}^n p(u_l)$. Next, distribute these sequences at random into $2^{n(R-4\epsilon)}$ bins. It is the bin index $m$ that we wish to send. This is accomplished by sending any $U^n(i)$ in bin $m$. For encoding, given the state $S_1^n$ and the message $m \in \{1, \ldots, 2^{nR}\}$, look in bin $m$ for a sequence $U^n(i)$ such that the $(U^n(i), S_1^n)$ pair is jointly typical. Send the associated jointly typical $X^n$. The decoder receives $Y^n$ according to the distribution $\prod_{l=1}^n p(y_l|x_l, s_{1,l}, s_{2,l})$, and observes $S_2^n$. The decoder looks for the unique sequence $U^n(k)$ such that $(U^n(k), Y^n, S_2^n)$ is strongly jointly typical and lets $\hat{W}$ be the index of the bin containing this $U^n(k)$.

There are three sources of potential error. Let $E_1$ be the event that, given $S_1^n$ and the message index $m$, there is no jointly typical $(U^n(i), S_1^n)$ in bin $m$. Note that because for a fixed $p(u|s_1)$, the capacity $C$ is a convex function of the distribution $p(x|u, s_1)$, we can assume that $p(x|u, s_1)$ is a deterministic function $f$, that is $p(x|u, s_1)$ takes values 0 or 1 only. Therefore, we assume that $x = f(u, s_1)$ if and only if $p(x|u, s_1) = 1$. Without loss of generality, we assume that message 1 is transmitted. Let $E_2$ be the event that $(U^n(1), Y^n, S_2^n)$ is not jointly typical, and $E_3$ be the event that $(U^n(k), Y^n, S_2^n)$ is jointly typical for some $k \neq 1$. The decoder will either decode incorrectly or declare an error in events $E_2$ and $E_3$.

We first analyze the probability of $E_1$. The probability that a pair $(U^n, S_1^n)$ is strongly jointly typical is greater than $(1-\epsilon)2^{-nI(U;S_1)-n\epsilon}$ for $n$ sufficiently large. There are a total of $2^{nI(U;Y,S_2)-2n\epsilon}$ $U^n(i)$'s, and $2^{nI(U;Y,S_2)-nI(U;S_1)-4n\epsilon}$ bins. Thus, the expected number of jointly typical codewords in a given bin is greater than $(1-\epsilon)2^{n\epsilon}$. Consequently, by a standard argument, $\Pr(E_1) \to 0$ as $n \to \infty$.

For the probability of $E_2$, we note by the Markov lemma that if $(U^n, X^n, S_1^n)$ is strongly jointly typical, then $(U^n, X^n, S_1^n, S_2^n, Y^n)$ will be strongly jointly typical with high probability. This shows that $\Pr\{E_2|E_1^c\}$ also goes to 0 as $n \to \infty$.

The third source of potential error is that some other $U^n$ is jointly typical with $(Y^n, S_2^n)$. But a $U^n$ being jointly typical with $(Y^n, S_2^n)$ has probability at most $2^{-nI(U;Y,S_2)+n\varepsilon}$. Since there are only $2^{nI(U;Y,S_2)-2n\epsilon} - 1$ other $U^n(i)$ sequences, we have

$$\Pr\{E_3\} \leq 2^{-nI(U;Y,S_2)+n\varepsilon} 2^{nI(U;Y,S_2)-2n\epsilon}$$

which tends to zero as $n \to \infty$.

This shows that all three error events are of arbitrarily small probability. By the union bound on these three probabilities of error, the average probability of error $P_e^n$ tends to zero. This concludes the proof of achievability.

We now prove the converse. We wish to show that $P_e^n \to 0$ implies $R \leq C$. This is equivalent to showing that there exists a distribution $p(s_1, s_2)p(u, x|s_1)p(y|s_1, s_2, x)$ on $(U, S_1, S_2, X, Y)$, with specified $p(s_1, s_2)p(y|s_1, s_2, x)$, such that

$$I(U; S_2, Y) - I(U; S_1) \geq R - \delta(P_e)$$

where $\delta(P_e) = -P_e \log P_e - (1 - P_e) \log(1 - P_e) + P_e \log |\mathcal{Y}|$.

First, define two auxiliary random variables as follows. Let

$$U_i = (W, Y^{i-1}, S_2^{i-1}, S_{1,i+1}^n) \quad \text{and} \quad V_i = (W, S_{1,i+1}^n),$$
$$\text{for } i = 1, 2, \ldots, n.$$

Thus,

$$U_i = (V_i, Y^{i-1}, S_2^{i-1})$$
$$V_{i-1} = (V_i, S_{1,i})$$
$$V_1 = U_1$$
$$V_n = W.$$

We will need the following.

*Lemma 1:*

$$-I(V_i; Y^i, S_2^i) + I(V_i; S_1^i) + I(V_{i-1}; Y^{i-1}, S_2^{i-1})$$
$$- I(V_{i-1}; S_1^{i-1}) + I(U_i; Y_i, S_{2,i})$$
$$- I(U_i; S_{1,i}) = I(Y^{i-1}; Y_i).$$

We postpone the proof of this lemma until after finishing the proof of the main theorem. Using the above lemma, we have the following inequalities:

$$I(Y^{i-1}; Y_i) \overset{(a)}{\geq} 0$$
$$I(V_{i-1}; Y^{i-1}, S_2^{i-1}) - I(V_{i-1}; S_1^{i-1}) + I(U_i; Y_i, S_{2,i})$$
$$- I(U_i; S_{1,i}) \overset{(b)}{\geq} I(V_i; Y^i, S_2^i) - I(V_i; S_1^i)$$
$$\sum_{i=1}^n [I(U_i; Y_i, S_{2,i}) - I(U_i; S_{1,i})]$$
$$\overset{(c)}{\geq} I(W; Y^n, S_2^n) - I(W; S_1^n)$$

where inequality (a) follows from nonnegativity of mutual information, inequality (b) follows from Lemma 1, and inequality (c) follows from summing over $i = 2$ to $n$. Then, letting $R_i =$

$I(U_i; Y_i, S_{2,i}) - I(U_i; S_{1,i})$, we have the following chain of equalities:

$$\sum_{i=1}^{n} R_i \geq I(W; Y^n, S_2^n) - I(W; S_1^n)$$
$$\stackrel{(d)}{=} I(W; Y^n, S_2^n)$$
$$\stackrel{(e)}{=} H(W) - H(W|Y^n, S_2^n)$$
$$\stackrel{(f)}{=} nR - H(W|Y^n, S_2^n)$$
$$\stackrel{(g)}{=} n(R - \delta(P_e))$$

where equality (d) follows from independence of $W$ and $S_1^n$, equality (e) follows from the definition of mutual information, equality (f) follows from the specified uniform distribution over $W$, and equality (g) follows from Fano's inequality and our definition of $\delta(P_e)$.

Now choosing $i^*$ to be the first index $i$ such that $R_i = \max_{1 \leq j \leq n} R_j$, we have

$$R_{i^*} \geq R - \delta(P_e).$$

Therefore, we have shown that

$$I(U_{i^*}; Y_{i^*}, S_{2,i^*}) - I(U_{i^*}; S_{1,i^*}) \geq R - \delta(P_e)$$

for the distribution $p(u_{i^*}, y_{i^*}, s_{1,i^*}, s_{2,i^*})$ induced by the code. Note that $\delta(P_e) \to 0$ as $P_e \to 0$. This concludes the converse except for the proof of Lemma 1.

We now prove Lemma 1. We follow the argument in [11]. Since

$$I(A; B) - I(A; B|C) = I(A; C) - I(A; C|B)$$

we have the following equalities:

$$I(V_i; S_{1,i}|Y^{i-1}, S_2^{i-1}) - I(V_i; S_{1,i})$$
$$= I(V_i; Y^{i-1}, S_2^{i-1}|S_{1,i}) - I(V_i; Y^{i-1}, S_2^{i-1})$$
$$I(V_i; Y^{i-1}, S_2^{i-1}) - I(V_i; Y^{i-1}, S_2^{i-1}|S_{1,i})$$
$$= I(V_i; S_{1,i}) - I(V_i; S_{1,i}|Y^{i-1}, S_2^{i-1}).$$

The following inequalities follow directly from information-theoretic identities, definitions of $U_i$ and $V_i$, and the fact that $(S_{1,i}, S_{2,i})$ are i.i.d.

$$I(V_i; Y^i, S_2^i) = I(V_i; Y^{i-1}, S_2^{i-1}) + I(V_i; Y_i, S_{2,i}|Y^{i-1}, S_2^{i-1})$$
$$I(V_i; S_1^i) = I(V_i; S_{1,i}) + I(V_i; S_1^{i-1}|S_{1,i})$$
$$I(V_{i-1}; S_1^{i-1}) = I(V_i; S_1^{i-1}|S_{1,i}) + I(S_{1,i}; S_1^{i-1})$$
$$= I(V_i; S_1^{i-1}|S_{1,i})$$
$$I(V_{i-1}; Y^{i-1}, S_2^{i-1}) = I(V_i, S_{1,i}; Y^{i-1}, S_2^{i-1})$$
$$= I(S_{1,i}; Y^{i-1}, S_2^{i-1})$$
$$+ I(V_i; Y^{i-1}, S_2^{i-1}|S_{1,i})$$
$$I(U_i; S_{1,i}) = I(Y^{i-1}, S_2^{i-1}; S_{1,i})$$
$$+ I(V_i; S_{1,i}|Y^{i-1}, S_2^{i-1})$$
$$I(U_i; Y_i, S_{2,i}) = I(Y^{i-1}, S_2^{i-1}; Y_i, S_{2,i})$$
$$+ I(V_i; Y_i, S_{2,i}|Y^{i-1}, S_2^{i-1}).$$

We now alternately attach positive and negative signs to the six equalities above and add them

$$I(V_i; Y^i, S_2^i) - I(V_i; S_1^i) + I(V_{i-1}; S_1^{i-1})$$
$$- I(V_{i-1}; Y^{i-1}, S_2^{i-1}) + I(U_i; S_{1,i}) - I(U_i; Y_i, S_{2,i})$$
$$= I(V_i; Y^{i-1}, S_2^{i-1}) + I(V_i; Y_i, S_{2,i}|Y^{i-1}, S_2^{i-1})$$
$$- I(V_i; S_{1,i}) - I(V_i; S_1^{i-1}|S_{1,i})$$
$$+ I(V_i; S_1^{i-1}|S_{1,i}) - I(S_{1,i}; Y^{i-1}, S_2^{i-1})$$
$$- I(V_i; Y^{i-1}, S_2^{i-1}|S_{1,i}) + I(Y^{i-1}, S_2^{i-1}; S_{1,i})$$
$$+ I(V_i; S_{1,i}|Y^{i-1}, S_2^{i-1}) - I(Y^{i-1}, S_2^{i-1}; Y_i, S_{2,i})$$
$$- I(V_i; Y_i, S_{2,i}|Y^{i-1}, S_2^{i-1}).$$

After cancellation and using the fact $I(Y^{i-1}; S_{2,i}|Y_i) = 0$, we obtain

$$-I(V_i; Y^i, S_2^i) + I(V_i; S_1^i) + I(V_{i-1}; Y^{i-1}, S_2^{i-1})$$
$$- I(V_{i-1}; S_1^{i-1}) + I(U_i; Y_i, S_{2,i}) - I(U_i; S_{1,i})$$
$$= I(Y^{i-1}; Y_i).$$

This proves the lemma and therefore concludes the converse.

## VIII. PROOF OF THEOREM 2

The proof in this section closely follows the proof in [19].

We prove Theorem 2 in two parts. We assume finite alphabets for $X$, $S_1$, $S_2$, $\hat{X}$, $U$. First, we show that

$$R(D) = \min_{p(u|x,s_1)p(\hat{x}|u,s_2)}[I(U; S_1, X) - I(U; S_2)]$$

is achievable. Given $p(s_1, s_2, x)$, prescribe

$$p(u|x, s_1) \quad \text{and} \quad p(\hat{x}|u, s_2).$$

This provides the joint distribution $p(x, s_1, s_2, u, \hat{x})$ and a distortion $D = Ed(X, \hat{X})$.

The overall idea is as follows. If we can arrange a coding $i(\cdot)$, $\hat{X}^n(\cdot)$ so that $(U^n, S_1^n, S_2^n, X^n, \hat{X}^n)$ appears to be distributed i.i.d. according to $p(s_1, s_2, x)$, then $2^{nI(U; X, S_1)}$ $U^n$'s will suffice to "cover" $\mathcal{X}^n \times \mathcal{S}_1^n$. Knowledge of $U^n$ and $S_2^n$ will then allow the reconstruction $\hat{X}^n$ to achieve distortion $D = Ed(X, \hat{X})$. The receiver has a list of $2^{nI(U; X, S_1)}$ possible $U^n(j)$'s, which he reduces by a factor of $2^{-nI(U; S_2)}$ by his knowledge of $S_2^n$ and by another factor of $2^{-nR}$ by observing the index of one of the $2^{nR}$ random bins into which the $U^n(j)$'s have been placed. Letting $R > I(U; X, S_1) - I(U; S_2)$ provides enough bins so that $U^n(k)$ is identified uniquely. Thus, $U^n$ is known at the decoder and distortion $D$ is achieved.

We first generate codewords as follows. Let $R_1 = I(U; X, S_1) + \epsilon$. Generate $2^{nR_1}$ i.i.d. codewords $U^n$ according to $\prod_{l=1}^{n} p(u_l)$ and index them $U^n(j)$, $j \in 2^{nR_1}$. Let

$$R = I(U; X, S_1) - I(U; S_2) + 3\epsilon.$$

Randomly assign the indexes of the codewords to one of $2^{nR}$ bins using a uniform distribution over the indices of the bins. Let $B(i)$ be the set of indices assigned to bin $i$.

Given a source sequence $X^n$ and state information $S_1^n$, the encoder looks for an index $m \in 2^{nR_1}$ such that $(U^n(m), X^n, S_1^n)$ is strongly jointly typical. If there is no such $U^n$, $m$ is set to 0. The encoder then sends the index $i$ such that $m \in B(i)$.

The decoder, given state information $S_2^n$ and bin index $i$, looks for a $U^n(k)$ such that $k \in B(i)$ and $(U^n(k), S_2^n)$ is strongly jointly typical. If there is such a $k$, the decoder selects any $\hat{X}^n$ such that $(U^n(k), S_2^n, \hat{X}^n)$ is strongly jointly typical. If there is no such $k$, or more than one such $k$, the decoder sets $\hat{X}^n$ to be an arbitrary sequence and incurs maximal distortion. The probability of this error is exponentially small with block length $n$.

We now argue that all the error events are of vanishing probability. First, by the law of large numbers, for sufficiently large $n$, the probability that $(X^n, S_1^n, S_2^n)$ is not strongly jointly typical is exponentially small.

If $(X^n, S_1^n)$ is strongly jointly typical, then the probability that there is no $m$ such that $(X^n, S_1^n, U^n(m))$ is strongly typical is exponentially small provided that $R_1 \geq I(U; X, S_1)$. This follows from the standard rate distortion argument that $2^{nR_1}$ random $U^n$'s "cover" $\mathcal{X}^n \times \mathcal{S}_1^n$.

By the Markov lemma on typicality, given that $(X^n, S_1^n, U^n(m))$ is strongly jointly typical, the probability that $(X^n, S_1^n, U^n(m), S_2^n)$ is strongly jointly typical is nearly 1 and thus the probability that $(U^n(m), S_2^n)$ is not strongly jointly typical is exponentially small.

Furthermore, the probability that there is another $U^n(k)$ in the same bin that is strongly jointly typical with $S_2^n$ is bounded by the number of codewords in the bin times the probability of joint typicality

$$2^{n(R_1-R)}2^{-n(I(U;S_2)-\varepsilon)}$$

which tends to zero because $R_1 - R < I(U; S_2) - \varepsilon$.

This shows that all the error events have asymptotically zero probability. Since index $k$ is decoded correctly, $(\hat{X}^n, S_2^n, U^n(k))$ is strongly jointly typical. Since both $(X^n, S_1^n, S_2^n)$ and $(U^n(k), S_1^n, X^n)$ are strongly jointly typical, $(X^n, S_2^n, U^n(k))$ is also strongly jointly typical. Therefore, the empirical joint distribution is close to the original distribution $p(x, s_1, s_2)p(u|x, s_1)$, and $(X^n, \hat{X}^n)$ will have a joint distribution that is close to the minimum distortion-achieving distribution. Thus, $Ed(X^n, \hat{X}^n)$ will be $\leq D$. This concludes the achievability part of the proof.

We now prove the converse in Theorem 2. Let the encoding function be $i: \mathcal{X}^n \times \mathcal{S}_1^n \to \{1, 2, \ldots, 2^{nR}\}$, and the decoding function be $\hat{X}^n: \mathcal{S}_2^n \times \{1, 2, \ldots, 2^{nR}\} \to \hat{\mathcal{X}}^n$. We need to show that

$$E\left[d\left(X^n, \hat{X}(S_2^n, i(X^n, S_1^n))\right)\right] \leq D$$

implies

$$R \geq R(D)$$
$$= \min_{p(u|x,s_1)p(\hat{x}|u,s_2), Ed(X,\hat{X}) \leq D} [I(U; S_1, X) - I(U; S_2)].$$

We first prove a lemma on the convexity of $R(D)$.

*Lemma 2:*

$$R(D) = \min_{p(u|x,s_1)p(\hat{x}|u,s_2), Ed(X,\hat{X}) \leq D} [I(U; S_1, X) - I(U; S_2)]$$

is a convex function of $D$.

*Proof:* Let $(U_1, \hat{X}_1)$ and $(U_2, \hat{X}_2)$ be random variables with distributions achieving the minimum rate for distortions $D_1$ and $D_2$, respectively. Let $Q$ be a random variable that is independent of $U_1, U_2, \hat{X}_1$ and $\hat{X}_2$, where $Q$ assumes value 1 with probability $\lambda$ and value 2 with probability $1-\lambda$. Now consider the rate distortion problem with $D$ as the distortion constraint. Because of the linearity of the distortion in the distribution $p(s_1, s_2, x)p(u|x, s_1, q)p(\hat{x}|u, s_2, q)p(q)$, we have

$$D = \lambda D_1 + (1-\lambda)D_2.$$

Let $\tilde{U} = (Q, U_Q)$. We have

$$R(D) = \min_{p(u|x,s_1)p(\hat{x}|u,s_2)}[I(U; X, S_1) - I(U; S_2)]$$
$$\leq I(\tilde{U}; X, S_1) - I(\tilde{U}; S_2)$$
$$= H(X, S_1) - H(X, S_1|\tilde{U}) - H(S_2) + H(S_2|\tilde{U})$$
$$= H(X, S_1) - \lambda H(X, S_1|U_1) - (1-\lambda)H(X, S_1|U_2)$$
$$\quad - H(S_2) + \lambda H(S_2|U_1) + (1-\lambda)H(S_2|U_2)$$
$$= \lambda[I(U_1; X, S_1) - I(U_1; S_2)]$$
$$\quad + (1-\lambda)[I(U_2; X, S_1) - I(U_2; S_2)]$$
$$= \lambda R(D_1) + (1-\lambda)R(D_2).$$

Therefore, $R(D) \leq \lambda R(D_1) + (1-\lambda)R(D_2)$. This proves the convexity of $R(D)$.

We now start the proof of the converse. We assume as given that $(X_i, S_{1,i}, S_{2,i})$ are i.i.d. $\sim p(x, s_1, s_2)$. We are given some rate $R$ encoding function $i: \mathcal{X}^n \times \mathcal{S}_1^n \to 2^{nR}$ and a decoding function $\hat{X}^n: \mathcal{S}_2^n \times 2^{nR} \to \hat{\mathcal{X}}^n$. Let $T = i(X^n, S_1^n)$, and let $\hat{X}^n = \hat{X}^n(S_2^n, T)$. The resulting distortion is

$$D = Ed(X^n, \hat{X}^n(i(X^n, S_1^n), S_2^n)) = E\left[\frac{1}{n}\sum_{i=1}^n d(X_i, \hat{X}_i)\right].$$

We wish to show that if this code has distortion $D$, then

$$R \geq R(D) = \min_{p(u|x,s_1)p(\hat{x}|u,s_2)}[I(U; S_1, X) - I(U; S_2)].$$

Define $U_i = (T, S_2^{i-1}, S_{2,i+1}^n)$ and

$$\hat{X}_i = \hat{X}_i(S_2^n, T) = \hat{X}_i(U_i, S_{2,i}).$$

Note that $\hat{X}_i$ is a function of $(U_i, S_{2,i})$ since $(S_2^n, T) = (U_i, S_{2,i})$. We apply standard information theoretic inequalities to obtain the following chain of inequalities:

$$nR \geq H(T)$$
$$\geq H(T|S_2^n)$$
$$\geq I(X^n, S_1^n; T|S_2^n)$$
$$= \sum_{i=1}^n I(X_i, S_{1,i}; T|S_2^n, X^{i-1}, S_1^{i-1})$$
$$= \sum_{i=1}^n H(X_i, S_{1,i}|S_2^n, X^{i-1}, S_1^{i-1})$$
$$\quad - H(X_i, S_{1,i}|T, S_2^n, S_1^{i-1}, X^{i-1})$$
$$\stackrel{(a)}{=} \sum_{i=1}^n H(X_i, S_{1,i}|S_{2,i})$$
$$\quad - H(X_i, S_{1,i}|T, S_2^n, S_1^{i-1}, X^{i-1})$$

$$\geq \sum_{i=1}^{n} H(X_i, S_{1,i}|S_{2,i}) - H(X_i, S_{1,i}|T, S_2^n)$$

$$= \sum_{i=1}^{n} H(X_i, S_{1,i}|S_{2,i}) - H(X_i, S_{1,i}|U_i, S_{2,i})$$

$$= \sum_{i=1}^{n} I(X_i, S_{1,i}; U_i|S_{2,i})$$

$$= \sum_{i=1}^{n} H(U_i|S_{2,i}) - H(U_i|X_i, S_{1,i}, S_{2,i})$$

$$\stackrel{(b)}{=} \sum_{i=1}^{n} H(U_i|S_{2,i}) - H(U_i|X_i, S_{1,i})$$

$$= \sum_{i=1}^{n} I(U_i; X_i, S_{1,i}) - I(U_i; S_{2,i})$$

$$\geq \sum_{i=1}^{n} R(E[d(X_i, \hat{X}_i(U_i, S_{2,i}))])$$

$$\stackrel{(c)}{\geq} nR\left(E\left[\frac{1}{n}\sum_{i=1}^{n} d(X_i, \hat{X}_i(U_i, S_{2,i}))\right]\right)$$

$$\geq nR(D)$$

where equality (a) follows from the fact that $(X_i, S_{1,i})$ is independent of past and future $(S_{1,j}, S_{2,j}, X_j)$ given $S_{2,i}$, equality (b) follows from the fact that $U_i \to (X_i, S_{1,i}) \to S_{2,i}$ forms a Markov chain, since

$$p(x^n, s_1^n, s_2^n) = \prod_{i=1}^{n} p(x_i, s_{1,i}, s_{2,i})$$

and, thus, $S_{2,i}$ is dependent on $T$ only through $(X_i, S_{1,i})$. Inequality (c) follows from the convexity of $R(D)$ and Jensen's inequality.

Therefore, $R \geq R(D)$, where $R(D)$ is as defined in Theorem 2. This concludes the converse.

## IX. CONCLUSION

The known special cases $C_{00}$, $C_{01}$, $C_{10}$, $C_{11}$ and $R_{00}$, $R_{01}$, $R_{10}$, $R_{11}$ of channel capacity and rate distortion with state information involve one-sided state information or two-sided but identical state information. We establish the general capacity

$$C_{S_1, S_2} = \max_{p(u,x|s_1)} [I(U; S_2, Y) - I(U; S_1)]$$

and rate distortion

$$R_{S_1, S_2}(D) = \min_{p(u|x,s_1)p(\hat{x}|u,s_2)} [I(U; S_1, X) - I(U; S_2)]$$

for two-sided state information. This includes the one-sided state information theorems as special cases, and makes the duality apparent.

## ACKNOWLEDGMENT

The authors had useful communications with J. K. Su, J. J. Eggers, and B. Girod. They would like to thank S. Shamai and the reviewers for their detailed feedback. They also thank Arak Sutivong and Young-Han Kim for many useful discussions.